\DocumentMetadata{}
\documentclass[sigconf]{acmart}

\AtBeginDocument{%
}

\copyrightyear{2025}
\acmYear{2025}
\setcopyright{rightsretained}
\acmConference[WSDM '25]{Proceedings of the Eighteenth ACM International Conference on Web Search and Data Mining}{March 10--14, 2025}{Hannover, Germany}
\acmBooktitle{Proceedings of the Eighteenth ACM International Conference on Web Search and Data Mining (WSDM '25), March 10--14, 2025, Hannover, Germany}
\acmDOI{10.1145/3701551.3703544}
\acmISBN{979-8-4007-1329-3/25/03}

\makeatletter
\gdef\@copyrightpermission{
  \begin{minipage}{0.2\columnwidth}
   \href{https://creativecommons.org/licenses/by/4.0/}{\includegraphics[width=0.90\textwidth]{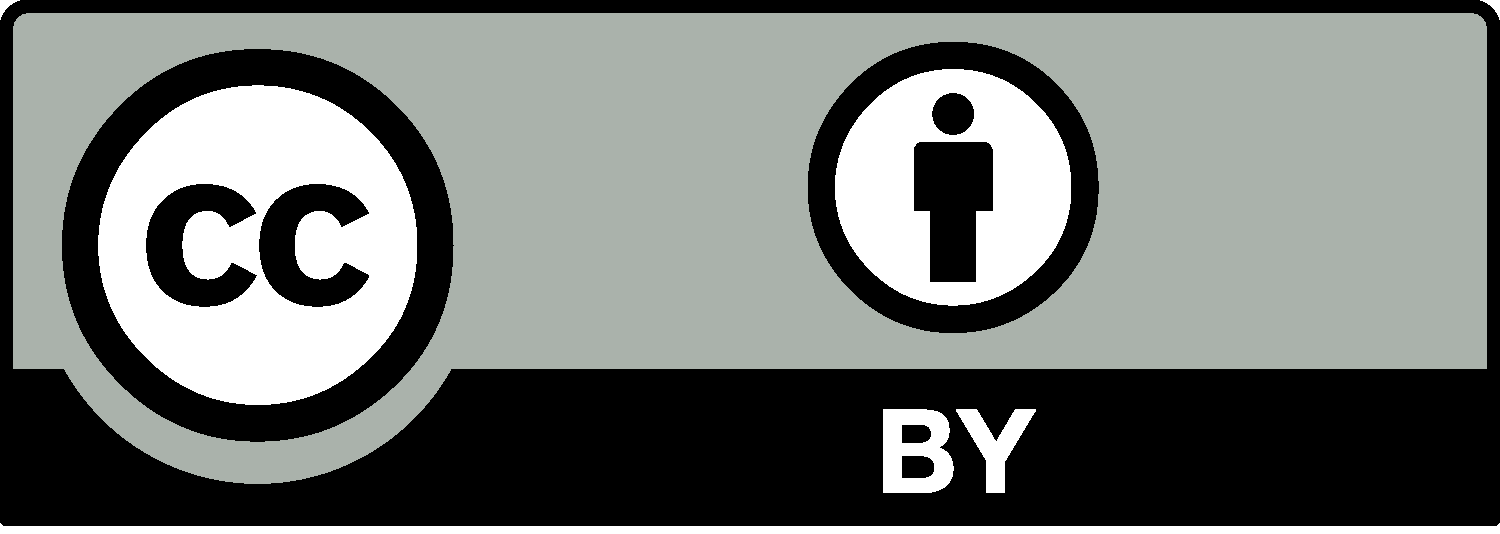}}
  \end{minipage}\hfill
  \begin{minipage}{0.8\columnwidth}
   \href{https://creativecommons.org/licenses/by/4.0/}{This work is licensed under a Creative Commons Attribution International 4.0 License.}
  \end{minipage}
  \vspace{5pt}
}
\makeatother

\usepackage[normalem]{ulem}
\useunder{\uline}{\ul}{}

\usepackage{hyperref}
\usepackage{pifont}
\usepackage{amsfonts, amsthm, array}
\usepackage[ruled,vlined,linesnumbered]{algorithm2e}
\usepackage{enumitem,multirow,graphicx,subcaption,multicol,lipsum,float,adjustbox}
\newlength{\textfloatsepsave}  
\usepackage{cleveref}
\usepackage{algorithmicx}
\usepackage[page]{appendix} 
\crefformat{section}{\S#2#1#3}
\crefformat{subsection}{\S#2#1#3}
\crefformat{subsubsection}{\S#2#1#3}
\usepackage{arydshln}
\usepackage{epstopdf}

\newcolumntype{P}{>{\raggedright\arraybackslash}m{0.98\linewidth}}
\newcolumntype{C}{>{\arraybackslash}m{1.\linewidth}}

\usepackage{makecell}

\begin{document}


\title{Improving Scientific Document Retrieval with~Concept~Coverage-based Query Set Generation}

\author{SeongKu Kang}
\affiliation{%
    \institution{Korea University}
    \city{Seoul}
    \country{South Korea}}
\email{seongkukang@korea.ac.kr}

\author{Bowen Jin}
\affiliation{
    \institution{University of Illinois at Urbana-Champaign}
    \city{Champaign}
    \country{United States}
}
\email{bowenj4@illinois.edu}

\author{Wonbin Kweon}
\affiliation{
    \institution{Pohang University of Science and Technology}
        \city{Pohang}
    \country{South Korea}
}
\email{kwb4453@postech.ac.kr}

\author{Yu Zhang}
\affiliation{
    \institution{Texas A\&M University}
    \city{College Station}
    \country{United States}
}
\email{yuzhang@tamu.edu}

\author{Dongha Lee}
\affiliation{
    \institution{Yonsei University}
        \city{Seoul}
    \country{South Korea}
}
\email{donalee@yonsei.ac.kr}

\author{Jiawei Han}
\affiliation{
    \institution{University of Illinois at Urbana-Champaign}
    \city{Champaign}
    \country{United States}
}
\email{hanj@illinois.edu}

\author{Hwanjo Yu}
\affiliation{
    \institution{Pohang University of Science and Technology}
    \city{Pohang}
    \country{South Korea}
}
\authornote{Corresponding author}
\email{hwanjoyu@postech.ac.kr}

\renewcommand{\shortauthors}{SeongKu Kang et al.}

\begin{abstract}
In specialized fields like the scientific domain, constructing large-scale human-annotated datasets poses a significant challenge due to the need for domain expertise.
Recent methods have employed large language models to generate synthetic queries, which serve as proxies for actual user queries.
However, they lack control over the content generated, often resulting in incomplete coverage of academic concepts in documents. 
We introduce \textbf{C}oncept \textbf{C}overage-based \textbf{Q}uery set \textbf{Gen}eration (\textbf{\proposed}) framework, designed to generate a set of queries with comprehensive coverage of the document's concepts.
A key distinction of \proposed is that it adaptively adjusts the generation process based on the previously generated queries.
We identify concepts not sufficiently covered by previous queries, and leverage them as conditions for subsequent query generation.
This approach guides each new query to complement the previous ones, aiding in a thorough understanding of the document.
Extensive experiments demonstrate that \proposed significantly enhances query quality and~retrieval~performance.

\end{abstract}

\newcommand{\proposed}{CCQGen\xspace}
\newcommand{\proposedtwo}{CSR\xspace}

\newcommand{\smallsection}[1]{{\vspace{0.03in} \noindent \bf {#1.}}}

\newcommand{\ctr}{{Contriever-MS}\xspace}
\newcommand{\specter}{{SPECTER-v2}\xspace}
\newcommand{\csfcube}{CSFCube\xspace}
\newcommand{\dorismae}{DORIS-MAE\xspace}

\maketitle

\section{Introduction}
Scientific document retrieval is a fundamental task that accelerates scientific innovations and access to technical solutions \cite{taxoindex}.
Recently, pre-trained language models (PLMs) have largely enhanced various ad-hoc searches \cite{CTR, DPR}. 
PLM-based retrievers are initially pre-trained on massive textual corpora to develop language understanding.
They are then fine-tuned using vast datasets of annotated query-document pairs, enabling the models to accurately assess the relevance between queries and documents.
However, in specialized domains like scientific document retrieval, constructing large-scale annotated datasets is challenging due to the need for domain expertise \cite{li2023sailer, ToTER, inpars}.
While there are a few general domain datasets (e.g., web search \cite{msmarco_data, NQ_data}), they often fail to generalize to specialized domains \cite{BEIR, inpars}.
This remains a major obstacle~for~applications.

\begin{figure}[t]
    \centering    
    \hspace{-0.2cm}
    \includegraphics[width=0.9\linewidth]{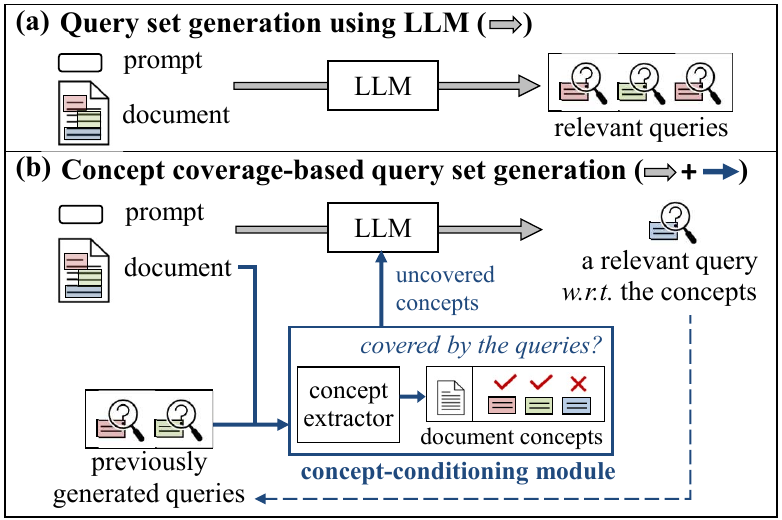}
    \caption{A conceptual comparison of (a) the existing approach for query set generation and (b) our concept coverage-based query set generation. Best viewed in color.}
    \label{fig:intro}
\end{figure}

Recently, large language models (LLMs) \cite{GPT3, FLAN, Tzero, Llama} have been actively utilized to generate synthetic data.
Given a document and a prompt including an instruction such as ``\textit{generate five relevant queries to the document}'' \cite{synthetic_apple_VA, dai2022promptagator}, LLMs generate synthetic queries for each document (Figure \ref{fig:intro}a).
The generated queries serve as proxies for actual user queries.
Recent developments in prompting schemes have largely improved the quality of these queries. 
\cite{inpars, dai2022promptagator} show that incorporating a few examples of actual query-document pairs in the prompt leads to the generation of queries with similar distributions (e.g., expression styles) to actual queries.
The state-of-the-art method \cite{pairwise_qgen} employs a \textit{pair-wise} generation that instructs LLMs to generate relevant queries first and then relatively less relevant ones. 
These less relevant queries serve as natural `hard negatives', further improving the efficacy of fine-tuning \cite{pairwise_qgen}.


\begin{table}[t]
    \caption{An example of synthetic queries. 
    The queries are generated sequentially from $q^1$ to $q^3$. 
    \proposed is applied after generating $q^1$.
    Repeated keywords are denoted in \textcolor{red}{\textbf{red}}, while newly covered concepts are denoted in \textcolor{blue}{\textbf{blue}}. 
    Details of the generation process are illustrated in Figure \ref{fig:method}.
    }\small
    \centering
    \renewcommand{\arraystretch}{0.75}
    \resizebox{1\linewidth}{!}{
    \begin{tabular}{C} \toprule
    \textbf{Document} \\ \midrule
     Automated music playlist generation is a specific form of music recommendation. 
     Collaborative filtering methods can be used to ... 
     However, the scarcity of thoroughly curated playlists and the bias towards popular songs ... we propose an alternative model based on a song-to-playlist classifier, ... while leveraging song features derived from audio, ... robust performance when recommending rare and out-of-set songs. ...
     \\ \midrule
     \textbf{Generated queries for the document} \\ \midrule
    \begin{enumerate}[leftmargin=*, after={\vspace*{-0.7\baselineskip}}]
         \item[$q^1$] How can \textcolor{black}{song-to-playlist classifiers} enhance \textcolor{black}{automated music playlist generation}?  
         \item[$q^{2}$] How can \textcolor{red}{automated playlist creation} be boosted through \textcolor{red}{song-to-playlist classification} and \textcolor{blue}{\textbf{feature exploitation}}?  $\,$ (\textit{w/o any condition})
         \item[$q^3$] How does \textcolor{red}{song-to-playlist classifier} differ from traditional \textcolor{blue}{\textbf{collaborative filtering}} for music recommendation? $\quad\,\,\,\,\,\,\,\,\,\,\,$ (\textit{w/o any condition})
    \end{enumerate}\vspace{-\topsep} \\ \hdashline
    \begin{enumerate}[leftmargin=*, after={\vspace*{-\baselineskip}}]  
      \item[$q^{2'}$] What techniques can be used to \textcolor{blue}{\textbf{overcome filter bubbles}} and \textcolor{blue}{\textbf{recommend out-of-set songs}}?  $\quad\quad\quad\quad\quad\quad\quad\quad\quad\quad\quad\quad$ (\textit{w/ \proposed})
     \item[$q^{3'}$] How to \textcolor{blue}{\textbf{leverage mel-spectrogram features}} to \textcolor{blue}{\textbf{mitigate the cold-start problem}} in \textcolor{black}{playlist recommendation}? $\quad\quad\quad\quad\quad\,$(\textit{w/ \proposed})
    \end{enumerate}\vspace{-\topsep} \\ \bottomrule
    \end{tabular}}
    \label{tab:intro}
    \vspace{0.3cm}
\end{table}

Though effective in generating plausible queries, the existing methods lack control over the content generated, which can lead to incomplete coverage of the academic concepts in a document.
Academic concepts refer to fundamental ideas, theories, and methodologies that form the contents of scientific documents. 
A scientific document typically explores various concepts.
For example, in Table~\ref{tab:intro}, the document addresses the primary task of music playlist recommendation, along with the design of classification-based models, solutions for popularity biases and data scarcity, and the utilization of audio features.
For a thorough understanding of the document, training queries should comprehensively cover these concepts.

%

However, in the absence of control over the content generated, the queries often repeatedly cover similar aspects of the document, showing high redundancy.
For example, in Table~\ref{tab:intro}, the generated queries ($q^2, q^3$) repeat keywords such as `automated playlist creation' and `song-to-playlist classification' already present in the previous query ($q^1$).
While these concepts are undoubtedly relevant to the document, such redundant queries cannot effectively bring new training signals.
Furthermore, the queries exhibit a particularly higher lexical overlap with the document, compared to the actual user queries (\cref{result:query_analysis}).
We observe that the queries tend to repeat only a few terms extracted from the document.
Given that users express the same concepts using various expressions in their queries, this limited term usage may not effectively simulate actual queries, reducing the efficacy of fine-tuning.
As a naive solution, one might consider adding more instructions to the prompt, such as ``\textit{use various terms and reduce redundancy among the queries}''.
However, this still lacks systematic control over the generation and fails to bring consistent improvements~(\cref{result:CCQGen});
the improved term usage often appears in common expressions (e.g., advance, enhance, and reinforce), not necessarily enhancing concept coverage.



We propose \textbf{C}oncept \textbf{C}overage-based \textbf{Q}uery set \textbf{Gen}eration (\proposed) framework to meet two desiderata for training queries: 
(1) The queries should cover complementary aspects, enabling comprehensive coverage of the document's concepts, and (2) The queries should articulate the concepts in various related terms, rather than merely echoing a few phrases from the document.
A key distinction of \proposed is that it adaptively adjusts the generation process based on the concept coverage of previously generated queries (Figure~\ref{fig:intro}b).
We introduce a \textit{concept extractor} to (1) identify the core concepts of each text and (2) uncover concept-related terms not explicitly mentioned in the document.
Using this information, we discern the concepts not sufficiently covered by previous queries, and leverage them as \textit{conditions} for the subsequent query generation.
Table~\ref{tab:intro} shows that the queries generated with \proposed ($q^{2'}, q^{3'}$) cover complementary concepts using more various~related~terms. 
Furthermore, we introduce new techniques to filter out low-quality queries and enhance retrieval accuracy using the obtained concept information.
Our primary contributions~are:
\begin{itemize}[leftmargin=*]\vspace{-\topsep}
    \item We show that existing query generation methods often fail to comprehensively cover academic concepts in documents, leading to suboptimal training and retrieval performance.
    
    \item We propose \proposed framework, which adaptively imposes conditions for subsequent generation based on the concept coverage.
    \proposed can be flexibly integrated with existing prompting schemes to enhance concept coverage of generated queries.

    \item We validate the effectiveness of \proposed by extensive experiments. 
    \proposed brings significant improvements in query quality and retrieval performance over existing prompting schemes.
\end{itemize}\vspace{-\topsep}

\section{Preliminaries}
\label{sec:preliminary}
\subsection{Fine-tuning Retrieval Model}
To perform retrieval on a new corpus, a PLM-based retriever is fine-tuned using a training set of annotated query-document pairs.
For each query $q$, the contrastive learning loss is typically applied:
\begin{equation}
    \mathcal{L} = -\log\frac{e^{s_{text}(q,\,\, d^+)}}{e^{s_{text}(q,\,\, d^+)} + \sum_{d^-} e^{s_{text}(q,\,\, d^-)}},
\end{equation}
where $d^+$ and $d^-$ denote the relevant and irrelevant documents. 
$s_{text}(q, d)$ represents the similarity score between the query and a document, computed by the retriever.
For effective fine-tuning, a substantial amount of training data is required. 
However, in specialized domains such as scientific document search, constructing vast human-annotated datasets is challenging due to the need for domain expertise, which remains an obstacle for~applications~\cite{li2023sailer, ToTER}.

\subsection{Prompt-based Query Generation}
\label{prelim:qgen}
Several attempts have been made to generate synthetic queries using LLMs. 
Recent advancements have centered on advancing prompting schemes to enhance the quality of these queries.
We summarize recent methods in terms of their prompting schemes.

\smallsection{Few-shot examples}
Several methods \cite{inpars, inpars2, dai2022promptagator, pairwise_qgen, label_condition_qgen, saad2023udapdr} incorporate a few examples of relevant query-document pairs in the prompt.
The prompt comprises the following components: $P = \{inst, (d_i, q_i)^k_{i=1}, d_t\}$, 
where $inst$ is the textual instruction\footnote{For example, ``\textit{Given a document, generate five search queries for which the document can be a perfect answer}''. 
The instructions vary slightly across methods, typically in terms of word choice.
In this work, we follow the instructions used in \cite{pairwise_qgen}.
}, 
$(d_i, q_i)^k_{i=1}$ denotes $k$ examples of the document and its relevant query, 
and $d_t$ is the new document we want to generate queries for.
By providing actual examples of the desired outputs, this technique effectively generates queries with distributions similar to actual queries (e.g., expression styles and lengths) \cite{dai2022promptagator}.
It is worth noting that this technique is also utilized in subsequent prompting schemes.

\smallsection{Label-conditioning} 
Relevance label $l$ (e.g., relevant and irrelevant) has been utilized to enhance query generation \cite{label_condition_qgen, saad2023udapdr, inpars}.
The prompt comprises $P = \{inst, (l_i, d_i, q_i)^k_{i=1}, (l_t, d_t)\}$, where $k$ label-document-query triplets are provided as examples.
$l_i$ represents the relevance label for the document $d_i$ and its associated query $q_i$.
To generate queries, the prompt takes the desired relevance label $l_t$ along with the document $d_t$.
This technique incorporates knowledge of different relevance, which aids in improving query quality and allows for generating both relevant and irrelevant queries \cite{label_condition_qgen}.

\smallsection{Pair-wise generation}
To further enhance the query quality, the state-of-the-art method \cite{pairwise_qgen} introduces a \textit{pair-wise} generation of relevant and irrelevant queries.
It instructs LLMs to first generate relevant queries and then generate relatively less relevant ones. 
The prompt comprises $P = \{inst, (d_i, q_i, q^-_i)^k_{i=1}, d_t\}$, where $q_i$ and $q^-_i$ denote relevant and irrelevant query for $d_i$, respectively.
The generation of irrelevant queries is conditioned on the previously generated relevant ones, allowing for generating thematically similar rather than completely unrelated queries.
These queries can serve as natural `hard negative' samples for training \cite{pairwise_qgen}.


\vspace{0.03in}
\textbf{Remarks.}
Though effective in generating plausible queries, there remains substantial room for improvement.
We observe that existing techniques often generate queries with limited coverage of the document's concepts.
That is, the queries frequently cover similar aspects of the document, exhibiting high redundancy and failing to add new training signals.
Furthermore, the queries show a high lexical overlap with the document, often repeating a few keywords from the document (\cref{result:query_analysis}).
Considering that the same concepts are expressed using diverse terms in actual user queries, merely repeating a few keywords may limit the efficacy of fine-tuning.


\begin{figure*}[t]
\centering
\includegraphics[width=1.0\textwidth]{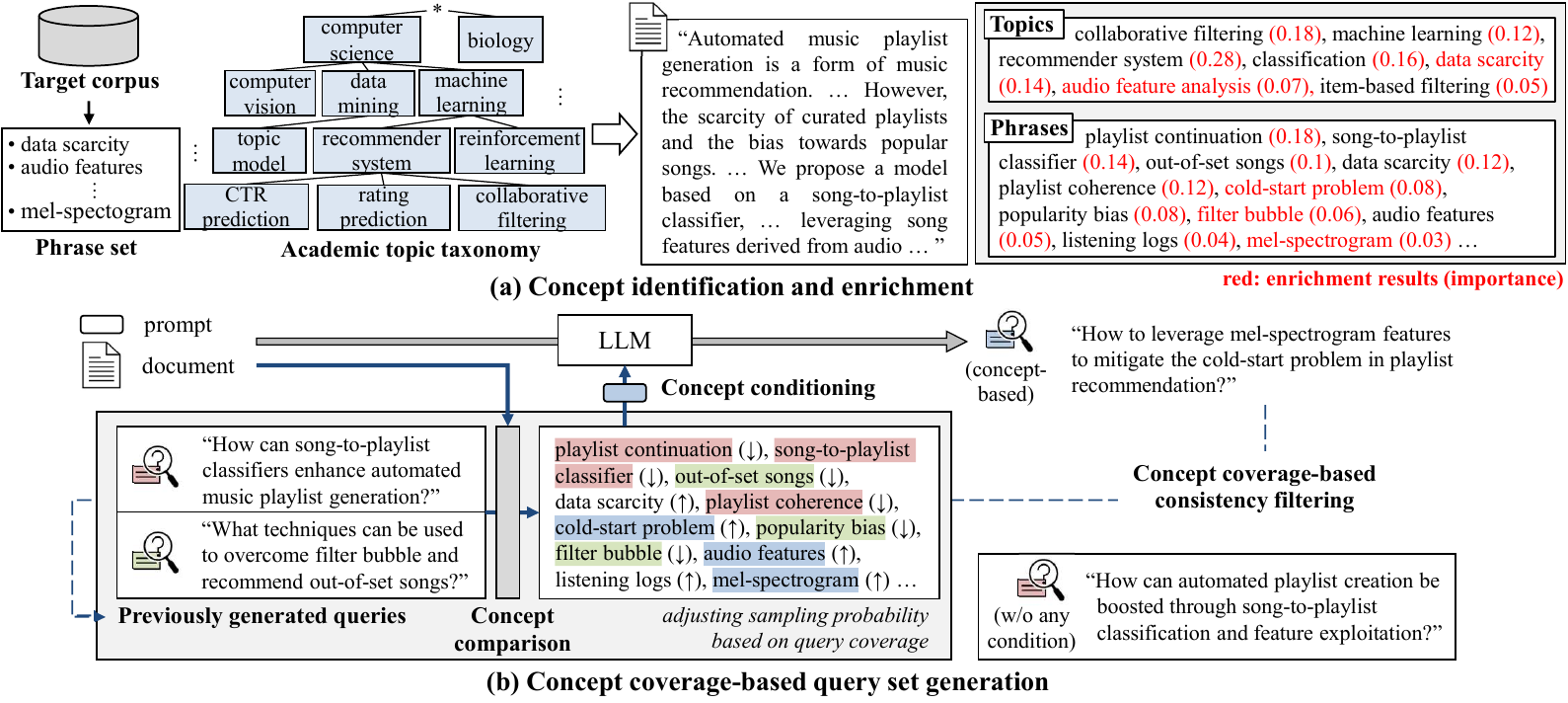}
\caption{The overview of Concept Coverage-based Query set Generation (\proposed) framework.
Best viewed in color.
}
\label{fig:method}
\vspace{-0.3cm}
\end{figure*}




\section{Methodology}
\label{sec:method}
We present \textbf{C}oncept \textbf{C}overage-based \textbf{Q}uery set \textbf{Gen}eration (\proposed) framework, designed to meet two desiderata of training queries:
(1) The concepts covered by queries should be complementary to each other, enabling a comprehensive coverage of the document’s concepts.
(2) The queries should articulate the document concepts in various related terms, rather than merely repeating phrases from the document.
\proposed consists of two major stages:
\begin{itemize}[leftmargin=*]\vspace{-\topsep}
    \item \textbf{Concept identification and enrichment (\cref{subsec:method_a}):}
        We first identify the core academic concepts of each document.
        Then, we enrich the identified concepts by assessing their importance and adding related concepts not explicitly mentioned in the document.
        This information serves as the basis for generating queries.
    \item \textbf{Concept coverage-based query generation (\cref{subsec:method_b}):}
        Given the previously generated queries $Q^{m-1}_d = \{q^1_d, ..., q^{m-1}_d\}$, we compare the concepts of the document $d$ with those covered by $Q^{m-1}_d$ to identify uncovered concepts.
        These uncovered concepts are then leveraged as conditions for generating the subsequent query $q^{m}_d$, allowing $q^{m}_d$ to cover complementary aspects of $Q^{m-1}_d$.
\end{itemize}\vspace{-\topsep}
Moreover, we propose a new technique, concept similarity-enhanced retrieval (\proposedtwo), that leverages the obtained concept information for \textbf{filtering out low-quality queries} and for \textbf{improving retrieval accuracy (\cref{subsub:method_filtering})}.
Figure \ref{fig:method} provides an overview of \proposed.


\subsection{Concept Identification and Enrichment}
\label{subsec:method_a}
To measure concept coverage, we first identify the core academic concepts of each document.
We represent the concepts using a combination of two different granularities: topic and phrase levels (Figure \ref{fig:method}a).
Topic level provides broader categorizations of research, such as `collaborative filtering' or `machine learning', while phrase level includes specific terms in the document, such as `playlist continuation' or `song-to-playlist classifier', complementarily revealing the document concepts.

A tempting way to obtain these topics and phrases is to simply instruct LLMs to find them in each document.
However, this approach has several limitations:
the results may contain concepts not covered by the document, and there is always a potential risk of hallucination.
As a solution, we propose a new approach that first constructs a candidate set, and then uses LLMs to pinpoint the most relevant ones from the given candidates, instead of directly generating them.
By doing so, the output space is restricted to the predefined candidate space, greatly reducing the risk of hallucinations while effectively leveraging the language-understanding capability of LLMs.

\subsubsection{\textbf{Core topics identification}}
\label{subsub:core_topic}
To identify the core topics of documents, we propose using an \textit{academic topic taxonomy} \cite{MAG_FS}. 
In the scientific domain, academic taxonomies are widely used for categorizing studies in various institutions and can be easily obtained from the web.\footnote{E.g., IEEE Taxonomy (\href{https://www.ieee.org/content/dam/ieee-org/ieee/web/org/pubs/ieee-taxonomy.pdf}{\color{blue} link}), ACM Computing Classification System (\href{https://dl.acm.org/ccs}{\color{blue} link}).}
A taxonomy refers to a hierarchical tree structure outlining academic topics (Figure \ref{fig:method}a).
Each node represents a topic, with child nodes corresponding to its sub-topics.
Leveraging taxonomy allows for exploiting domain knowledge of topic hierarchy and reflecting researchers' tendency to~classify~studies.

\smallsection{Candidate set construction}
One challenge in finding candidate topics is that the taxonomy obtained from the web is often very large and contains many irrelevant topics.
To effectively narrow down the candidates, we employ a \textit{top-down traversal} technique that \textit{recursively visits} the child nodes with the highest similarities at each level.
For each document, we start from the root node and compute its similarity to each child node.
We then visit child nodes with the highest similarities.\footnote{We visit multiple child nodes and create multiple paths, as a document usually covers various topics. For a node at level $l$, we visit $l+2$ nodes to reflect the increasing number of nodes at deeper levels of the taxonomy. The root~node~is~level~$0$.}
This process recurs until every path reaches leaf nodes, and \textit{all visited nodes} are regarded as candidates.

The document-topic similarity ${s}(d, c)$ can be defined in various ways.
As a topic encompasses its subtopics, we collectively consider the subtopic information for each topic node.
Let $\mathcal{N}_c$ denote the set of nodes in the sub-tree having $c$ as a root node.
We compute the similarity as: ${s}(d, c) = \frac{1}{|\mathcal{N}_c|}\sum_{j \in \mathcal{N}_c} \operatorname{cos}(\mathbf{e}_{d}, \mathbf{e}_{j})$, 
where $\mathbf{e}_d$ and $\mathbf{e}_j$ denote representations from PLM for a document $d$ and the topic name of node $j$, respectively.\footnote{We use BERT with mean pooling as the simplest choice.}

\smallsection{Core topic selection}
We instruct LLMs to select the most relevant topics from the candidates.
An example of an input prompt is:
\begin{table}[h]
\small
    \centering
    \resizebox{1.0\linewidth}{!}{
    \begin{tabular}{|C|}
    \hline
    You will receive a document along with a set of candidate topics. Your task is to select the topics that best align with the core theme of the document. 
    Exclude topics that are too broad or less relevant.
    You may list up to [$k^t$] topics, using only the topic names in the candidate set. \textbf{Document}:~[\textsc{Document}],~\textbf{Candidate~topic~set}:~[\textsc{Candidates}]\\ \hline 
    \end{tabular}}
    \vspace{-0.3cm}
\end{table}

In this work, we set $k^t=10$.
For each document $d$, we obtain core topics as $\mathbf{y}^t_d \in \{0,1\}^{|\mathcal{T}|}$, where $y^t_{di}=1$ indicates $i$ is a core topic of $d$, otherwise $0$.
$\mathcal{T}$ denotes the topic set obtained~from~the~taxonomy.


%




%

\subsubsection{\textbf{Core phrases identification}}
\label{method:core_phrase}
From each document, we identify core phrases used to describe its concepts.
These phrases offer fine-grained details not captured at the topic level.
We note that not all phrases in the document are equally important.
Core phrases should describe concepts strongly relevant to the document but \textit{not frequently covered} by other documents with similar topics.
For example, among documents about `recommender system' topic, the phrase `user-item interaction' is very commonly used, and less likely to represent the most important concepts~of~the~document.

\smallsection{Candidate set construction}
Given the phrase set $\mathcal{P}$ of the corpus\footnote{The phrase set is obtained using an off-the-shelf phrase mining tool \cite{autophrase}.}, we measure the distinctiveness of phrase $p$ in document $d$.
Inspired by recent phrase mining methods \cite{tao2016multi, lee2022taxocom}, we compute the distinctiveness as: $\exp(\operatorname{BM25}(p, d))/\,(1 + \sum_{d'\in\mathcal{D}_{d}}\exp(\operatorname{BM25}(p, d')))$.
This quantifies the relative relevance of $p$ to the document $d$ compared to other topically similar documents $\mathcal{D}_{d}$. 
$\mathcal{D}_{d}$ is simply retrieved using Jaccard similarity of core topics $\mathbf{y}^t_d$.
We set $|\mathcal{D}_{d}|=100$.
We select phrases with top-20\% distinctiveness score~as~candidates. 

\smallsection{Core phrase selection}
We instruct LLMs to select the most relevant phrases (up to $k^p$ phrases) from the candidates, using the same instruction format used for the topic selection.
We set $k^p=15$.
The core phrases are denoted by $\mathbf{y}^p_d \in \{0,1\}^{|\mathcal{P}|}$, where $y^p_{dj}=1$ indicates $j$ is a core phrase of $d$, otherwise $0$.

\subsubsection{\textbf{Enriching concept information}}
\label{method:enrich}
We have identified core topics and phrases representing each document's concepts.
We further enrich this information by (1) measuring their relative importance, and (2) incorporating strongly related concepts (i.e., topics and phrases) not explicitly revealed in the document.
This enriched information serves as the basis for generating queries.

\vspace{0.02in} \noindent
\textbf{Concept extractor.}
We employ a small model called a \textit{concept extractor}.
For a document $d$, the model is trained to predict its core topics $\mathbf{y}^{t}_{d}$ and phrases $\mathbf{y}^{p}_{d}$ from the PLM representation $\mathbf{e}_d$.
We formulate this as a two-level classification task: topic and~phrase~levels.


Topics and phrases represent concepts at different levels of granularity, and learning one task can aid the other by providing a complementary perspective.
To exploit their complementarity, we employ a multi-task learning model with two heads \cite{mmoe}.
Each head has a Softmax output layer, producing probabilities for topics $\hat{\mathbf{y}}^{t}_{d}$ and phrases $\hat{\mathbf{y}}^{p}_{d}$, respectively.
The cross-entropy loss is then applied for classification learning: $-\sum_{i=1}^{|\mathcal{T}|} y^t_{di} \log \hat{y}^{t}_{di} - \sum_{j=1}^{|\mathcal{P}|} y^p_{dj} \log \hat{y}^{p}_{dj}$.

\smallsection{Concept enrichment}
Using the trained concept extractor, we compute $\hat{\mathbf{y}}^{t}_{d}$ and $\hat{\mathbf{y}}^{p}_{d}$, which reveal their importance in describing the document's concepts.
Also, we identify strongly related topics and phrases that are expressed differently or not explicitly mentioned, by incorporating those with the highest prediction probabilities.
For example, in Figure \ref{fig:method}, we identify phrases `cold-start problem', `filter bubble', and `mel-spectrogram', which are strongly relevant to the document's concepts but not explicitly mentioned, along with their detailed importance.
These phrases are used to aid in articulating the document's concepts in various related terms.


We obtain $k^{t'}$ enriched topics and $k^{p'}$ enriched phrases for each document with their importance from $\hat{\mathbf{y}}^{t}_{d}$ and $\hat{\mathbf{y}}^{p}_{d}$.
We set the probabilities for the remaining topics and phrases as $0$, and normalize the probabilities for selected topics and phrases, denoted by $\bar{\textbf{y}}^t_d$ and $\bar{\textbf{y}}^p_d$.


\subsection{Concept Coverage-based Query Generation}
\label{subsec:method_b}
We present how we generate a set of queries that comprehensively cover the various concepts of a document.
We first identify concepts insufficiently covered by the previously generated queries (\cref{subsub:method_sampling}) and leverage them as conditions for subsequent generation (\cref{subsub:method_condition}).
Then, a filtering step is applied to ensure the query quality~(\cref{subsub:method_filtering}).

This process is repeated until a predefined number ($M$) of queries per document is achieved.
$M$ is empirically determined, considering available training resources such as GPU memory and training time.
For the first query of each document, we impose no conditions, thus it is identical to the results obtained from existing methods.

\subsubsection{\textbf{Concept sampling based on query coverage}}
\label{subsub:method_sampling}
The enriched information $\bar{\textbf{y}}_d$ reveals the core concepts and their importance within the document.
Our key idea is to generate queries that align with this distribution to ensure comprehensive coverage of the document's concepts.
Let $Q^{m-1}_d = \{q^1_d, ..., q^{m-1}_d\}$ denote the previously generated queries.
Using the concept extractor, which is trained to predict core concepts from the text, we identify the concepts covered by the queries, i.e., $\bar{\textbf{y}}^t_Q$ and $\bar{\textbf{y}}^p_Q$.
We use the concatenation of queries as input, denoted as $Q$.
A high value in $\bar{\textbf{y}}_d$ coupled with a low value in $\bar{\textbf{y}}_Q$ indicates that the existing queries do not sufficiently cover the corresponding concepts.

Based on the concept coverage information, we identify concepts that need to be more emphasized in the subsequently generated query.
We opt to leverage phrases as \textit{explicit} conditions for generation, as topics reveal concepts at a broad level, making them less effective for explicit inclusion in the query.
Note that topics are \textit{implicitly} reflected in identifying and enriching core phrases. 
We define a probability distribution to identify less~covered~concepts~as:
\begin{equation}
    \boldsymbol{\pi} = \operatorname{normalize}(\,\max(\bar{\textbf{y}}^p_d - \bar{\textbf{y}}^p_Q, \,\epsilon)\,)
\end{equation}
We set $\epsilon = 10^{-3}$ as a minimal value to the core phrases for numerical stability.
We sample $\lfloor \frac{k^{p'}}{M} \rfloor$ different phrases from $\operatorname{Multinomial}(\boldsymbol{\pi})$, where $M$ is the total number of queries per document.
Note that $\bar{\textbf{y}}^p_Q$ is dynamically adjusted during the construction of the~query~set.


\subsubsection{\textbf{Concept conditioning for query generation}}
\label{subsub:method_condition}
The sampled phrases are leveraged as conditions for generating the next query $q^m_d$.
There have been active studies to control the generation of LLMs for various tasks. 
Recent methods \cite{control_gen, outline_condition} have specified conditions for the desired outputs, such as sentiment, keywords, and an outline, directly in the prompts.
Following these studies, we impose a condition by adding a simple textual instruction $C$: ``\textit{Generate a relevant query based on the following keywords}: [\textsc{Sampled phrases}]''.
While more sophisticated instruction could be employed, we obtained satisfactory results with~our~choice.

The final prompt is constructed as $[P; C]$, where $P$ is an existing prompting scheme discussed in \cref{prelim:qgen}.
This integration allows us to inherit the benefits of existing techniques (e.g., few-shot examples), while generating queries that comprehensively cover the document's concepts.
For example, in Figure \ref{fig:method}, $C$ includes phrases like `cold-start problem' and `audio features', which are not well covered by the previous queries.
Based on this concept condition, we guide LLMs to generate a query that covers complementary aspects to the previous ones.
It is important to note that $C$ adds an \textit{additional condition} for $P$; the query is still about playlist recommendation, the main task of the document.

\subsubsection{\textbf{Concept coverage-based consistency filtering}}
\label{subsub:method_filtering}
After generating a query, we apply a filtering step to ensure its quality.
A critical criterion for this process is \textit{round-trip consistency} \cite{alberti2019synthetic}; a query should be answerable by the document from which it was generated.
Existing work \cite{dai2022promptagator, label_condition_qgen} employs a retriever to assess this consistency.
Given a generated pair $(q_d, d)$, the retriever retrieves documents for $q_d$. 
Then, $q_d$ is retained only if $d$ ranks within the top-$N$ results.
The accuracy of the retriever is crucial in this step;
a weak retriever may fail to filter out low-quality queries and also only retain queries that are \textit{too easy} (e.g., high lexical overlap with the document), thereby limiting the effectiveness of training.

We note that relying on the existing retriever is insufficient for measuring relevance.
While it is effective at capturing similarities of surface texts, the retriever often fails to match underlying concepts.
For example, in Figure \ref{fig:method}, the generated query includes phrases `cold-start problem' and `mel-spectrogram', which are highly pertinent to `data scarcity' and `audio features' discussed in the document.
Nevertheless, as these phrases are not directly used in the document, the retriever struggles to assess the relevance and ranks the document low. 
Consequently, the query is considered unreliable and removed during the filtering process.

\smallsection{Concept similarity-enhanced retrieval (CSR)}
We propose a simple and effective technique to enhance retrieval by using concept information.
For relevance prediction, we consider both textual similarity from the retriever $s_{text}(q,d)$, and concept similarity $s_{concept}(q,d)$.
We measure concept similarity using core phrase distributions, i.e., $s_{concept}(q,d) = sim(\bar{\mathbf{y}}^p_q, \,\bar{\mathbf{y}}^p_d)$, which reveals related concepts at a fine-grained level.\footnote{Here, we compute the similarity for top-10\% phrases (instead of $k^{p'}$) to consider concepts having a certain degree of relevance.
We also tried using core topics. However, it proved less effective as topics reveal concepts only~at~a~broad~level.
}
$sim(\cdot, \cdot)$ is the similarity function, for which use inner-product.
The relevance score~is~defined~as:
\begin{equation}
\begin{aligned}
rel_{CSR}(q,d) = f(s_{text}(q,d), \,s_{concept}(q,d)),
\end{aligned}
\end{equation}
where $f(\cdot, \cdot)$ is a function that combines the two scores. 
We use a simple addition after rescaling them via z-score normalization.
We denote this technique as Concept Similarity-enhanced Retrieval~(\proposedtwo).

For \textbf{filtering process}, we assess the round-trip consistency using \proposedtwo.
By directly matching underlying concepts not apparent from the surface text, we can more accurately measure relevance and distinguish low-quality queries.
Additionally, for \textbf{search with test queries} (i.e., after fine-tuning using the generated data), \proposedtwo can be used as a supplementary technique to further enhance retrieval.
It helps to understand test queries, which contain highly limited contexts and jargon not included in the training queries, by associating them with pre-organized~concept~information.


\section{Experiments}
\smallsection{\textbf{Datasets}}
We conduct a thorough review of the literature to find retrieval datasets in the scientific domain, specifically those where relevance has been assessed by skilled experts or annotators.
We select two recently published datasets: \textbf{\csfcube} \cite{CSFCube} and \textbf{\dorismae} \cite{DORISMAE}.
They offer test query collections annotated by human experts and LLMs, respectively, and embody two real-world search scenarios: query-by-example and human-written queries.
For both datasets, we conduct retrieval from the entire corpus, including all candidate documents.
\csfcube dataset consists of 50 test queries, with about 120 candidates per query drawn from approximately 800,000 papers in the S2ORC corpus \cite{lo2020s2orc}. 
\dorismae dataset consists of 165,144 test queries, with candidates drawn similarly to \csfcube.
We consider annotation scores above `2', which indicate documents are `nearly identical or similar' (\csfcube) and `directly answer all key components' (\dorismae), as relevant.
Note that training queries are not provided in both datasets.

\smallsection{\textbf{Academic topic taxonomy}}
We utilize the field of study taxonomy from Microsoft Academic \cite{MAG_FS}, which contains $431,416$ nodes with a maximum depth of $4$.
After the concept identification step (\cref{subsec:method_a}), we obtain $1,164$ topics and $18,440$ phrases for \csfcube, and $1,498$ topics and $34,311$ phrases for \dorismae.

\smallsection{\textbf{Metrics}}
Following \cite{mackie2023generative, ToTER}, we employ Recall@$K$ (R@$K$) for a large retrieval size ($K$), and NDCG@$K$ (N@$K$) and MAP@$K$ (M@K) for a smaller $K$ ($\leq 20$).
Recall@$K$ measures the proportion of relevant documents in the top $K$ results, while NDCG@$K$ and MAP@$K$ assign higher weights to relevant documents at higher~ranks.

\smallsection{\textbf{Backbone retrievers}}
We employ two representative models: 
(1) \textbf{\ctr} \cite{CTR} is a widely used retriever fine-tuned using vast labeled data from general domains (i.e., MS MARCO).
(2) \textbf{\specter} \cite{SPECTER2} is a PLM specifically developed for the scientific domain. It is trained using metadata (e.g., citation relations) of scientific papers. 
For both models, we use public checkpoints: \texttt{facebook/contriever-msmarco} and  \texttt{allenai/specter2\_base}.

\smallsection{\textbf{Baselines}}
We compare various query generation methods.
For all LLM-based methods, we use \texttt{gpt-3.5-turbo-0125}.
Additionally, we explore the results with a smaller LLM (\texttt{Llama-3-8B}) in \cref{result:Llama}.
For each document, we generate \textbf{five} relevant queries~\cite{BEIR}.
\begin{itemize}[leftmargin=*]\vspace{-0.7\topsep}
    \item \textbf{GenQ} \cite{BEIR} employs a specialized query generation model, trained with massive document-query pairs from the general domains.
    We use T5-base, trained using approximately $500,000$ pairs from MS MARCO dataset \cite{nogueira2019doc2query}: \texttt{BeIR/query-gen-msmarco-t5-base-v1}.
\end{itemize}
\noindent
\proposed can be flexibly integrated with existing LLM-based methods to enhance the concept coverage of the generated queries.
We apply \proposed to two recent approaches,  discussed~in~\cref{prelim:qgen}.
\begin{itemize}[leftmargin=*]\vspace{-0.7\topsep}
    \item \textbf{Promptgator} \cite{dai2022promptagator} is a recent LLM-based query generation method that leverages \textbf{few-shot examples} within the prompt. 
    

    \item \textbf{Pair-wise generation} \cite{pairwise_qgen} is the state-of-the-art method that generates relevant and irrelevant queries in a \textbf{pair-wise}~manner.
\end{itemize}
Additionally, we devise a new competitor that adds more instruction in the prompt to enhance the quality of queries:
\textbf{Promptgator\_{diverse}} is a variant of Promptgator, where we add the instruction ``\textit{use various terms and reduce redundancy among the~queries}''.

\smallsection{\textbf{Implementation details}}
We conduct all experiments using 4 NVIDIA RTX A5000 GPUs, 512 GB memory, and a single Intel Xeon Gold 6226R processor. 
For fine-tuning, we use top-50 BM25 hard negatives for each query \cite{formal2022distillation}.
We use 10\% of training data as a validation set. 
The learning rate is set to $10^{-6}$ for \ctr and $10^{-7}$ for \specter, after searching among $\{10^{-7}, 10^{-6}, ..., 10^{-3}\}$.
We set the batch size as $64$ and the weight decay as $10^{-4}$.
We report the average performance over five independent runs.
For all methods, we generate five synthetic queries for each document ($M=5$).
For the few-shot examples in the prompt, we randomly select five annotated examples, which are then excluded in the evaluation process \cite{dai2022promptagator}.
We follow the textual instruction used in \cite{pairwise_qgen}.
For other baseline-specific setups, we adhere to the configurations described in the original papers.
For the concept extractor, we employ a multi-gate mixture of expert architecture \cite{mmoe}, designed for multi-task learning.
We use three experts, each being a two-layer MLP.
For the consistency filtering, we set $N=5$.
We set the number of enriched topics and phrases to $k^{t'}=15$ and $k^{p'}=20$,~respectively.


\begin{table*}[t]
\caption{Retrieval performance comparison after fine-tuning with the generated queries. \textcolor{red}{Red} color denotes results that fail to show improvements over no Fine-Tune. $^{\dagger}$ and * indicate a statistically significant difference ($p<0.05$) from no Fine-Tune (one-sample t-test) and the applied query generation method (paired t-test), respectively.}
\centering
\renewcommand{\arraystretch}{0.95}
\resizebox{\linewidth}{!}{
\begin{tabular}{cl llllll  llllll}\toprule
& \multicolumn{1}{c}{\multirow{2}{*}{\textbf{Query generation}}} & \multicolumn{6}{c}{\textbf{CSFCube}} & \multicolumn{6}{c}{\textbf{DORIS-MAE}} \\ \cmidrule(lr){3-8} \cmidrule(lr){9-14}
 &  & \textbf{N@10} & \textbf{N@20} & \textbf{M@10} & \textbf{M@20} & \textbf{R@50} & \textbf{R@100} & \textbf{N@10} & \textbf{N@20} & \textbf{M@10} & \textbf{M@20} & \textbf{R@50} & \textbf{R@100} \\ \midrule
\parbox[t]{2mm}{\multirow{7}{*}{\rotatebox[origin=c]{90}{\ctr}}} & no Fine-Tune & 0.3313 & 0.3604 & 0.1525 & 0.1937 & 0.5783 & 0.7136 & 0.2603 & 0.2707 & 0.1177 & 0.1422 & 0.4509 & 0.5877 \\
 & GenQ & 0.3401 & \textcolor{red}{0.3495} & \textcolor{red}{0.1476} & \textcolor{red}{0.1841} & \textcolor{red}{0.5571} & \textcolor{red}{0.6843} & \textcolor{red}{0.2496} & \textcolor{red}{0.2647} & \textcolor{red}{0.1152} & \textcolor{red}{0.1396} & 0.4598 & \textcolor{red}{0.5805} \\
 & Promptgator\_{diverse} & 0.3539 & 0.3771 & 0.1606 & 0.2029 & 0.5950 & \textcolor{red}{0.7132} & \textcolor{red}{0.2461} & \textcolor{red}{0.2690} & \textcolor{red}{0.1143} & \textcolor{red}{0.1406} & 0.4645 & 0.5951 \\ \cmidrule(lr){2-14}
 & Promptgator & 0.3441 & 0.3670 & 0.1538 & 0.1974 & 0.5928 & 0.7298 & \textcolor{red}{0.2526} & 0.2724 & \textcolor{red}{0.1161} & \textcolor{red}{0.1418} & 0.4718 & 0.5961 \\
 & w/ CCQGen (ours) & \textbf{0.3605}$^{\dagger}$ & \textbf{0.3991}$^{\dagger}$$^*$ & \textbf{0.1614}$^{\dagger}$ & \textbf{0.2194}$^{\dagger}$$^*$ & \textbf{0.6333}$^{\dagger}$$^*$ & \textbf{0.7467}$^{\dagger}$ & \textbf{0.2697}$^*$ & \textbf{0.2883}$^{\dagger}$$^*$ & \textbf{0.1267}$^{\dagger}$$^*$ & \textbf{0.1536}$^{\dagger}$$^*$ & \textbf{0.4983}$^{\dagger}$$^*$ & \textbf{0.6327}$^{\dagger}$$^*$ \\ \cmidrule(lr){2-14}
 & Pair-wise generation & 0.3418 & 0.3686 & \textcolor{red}{0.1522} & 0.1971 & 0.5961 & 0.7225 & \textcolor{red}{0.2541} & 0.2753 & \textcolor{red}{0.1177} & 0.1445 & 0.4809 & 0.5947 \\
 & w/ CCQGen (ours) & \textbf{0.3670}$^{\dagger}$$^*$ & \textbf{0.4063}$^{\dagger}$$^*$ & \textbf{0.1656}$^{\dagger}$$^*$ & \textbf{0.2228}$^{\dagger}$$^*$ & \textbf{0.6362}$^{\dagger}$$^*$ & \textbf{0.7526}$^{\dagger}$$^*$ & \textbf{0.2783}$^{\dagger}$$^*$ & \textbf{0.2943}$^{\dagger}$$^*$ & \textbf{0.1308}$^{\dagger}$$^*$ & \textbf{0.1577}$^{\dagger}$$^*$ & \textbf{0.5089}$^{\dagger}$$^*$ & \textbf{0.6331}$^{\dagger}$$^*$ \\ \midrule
\parbox[t]{2mm}{\multirow{7}{*}{\rotatebox[origin=c]{90}{SPECTER-v2}}} & no Fine-Tune & 0.3503 & 0.3579 & 0.1615 & 0.2043 & 0.5341 & 0.6859 & 0.2121 & 0.2283 & 0.0942 & 0.1147 & 0.4182 & 0.5441 \\
 & GenQ & 0.3658 & 0.3659 & 0.1699 & 0.2083 & 0.5541 & \textcolor{red}{0.6836} & 0.2338 & 0.2525 & 0.1045 & 0.1287 & 0.4412 & 0.5613 \\
 & Promptgator\_diverse & 0.3672 & 0.3801 & 0.1721 & 0.2157 & 0.5687 & 0.6972 & 0.2469 & 0.2733 & 0.1121 & 0.1401 & 0.4843 & 0.6102 \\ \cmidrule(lr){2-14}
 & Promptgator & 0.3766 & 0.3886 & 0.1790 & 0.2245 & 0.5715 & 0.6962 & 0.2479 & 0.2713 & 0.1131 & 0.1398 & 0.4851 & 0.6064 \\
 & w/ CCQGen (ours) & \textbf{0.4105}$^{\dagger}$$^*$ & \textbf{0.4176}$^{\dagger}$$^*$ & \textbf{0.2085}$^{\dagger}$$^*$ & \textbf{0.2549}$^{\dagger}$$^*$ & \textbf{0.5886}$^{\dagger}$ & \textbf{0.7355}$^{\dagger}$$^*$ & \textbf{0.2634}$^{\dagger}$$^*$ & \textbf{0.2891}$^{\dagger}$$^*$ & \textbf{0.1226}$^{\dagger}$ & \textbf{0.1520}$^{\dagger}$$^*$ & \textbf{0.4988}$^{\dagger}$ & \textbf{0.6265}$^{\dagger}$ \\ \cmidrule(lr){2-14}
 & Pair-wise generation & 0.3870 & 0.3999 & 0.1966 & 0.2423 & 0.5722 & 0.6972 & 0.2523 & 0.2782 & 0.1163 & 0.1442 & 0.4885 & 0.6148 \\
 & w/ CCQGen (ours) & \textbf{0.4031}$^{\dagger}$$^*$ & \textbf{0.4150}$^{\dagger}$ & \textbf{0.2040}$^{\dagger}$ & \textbf{0.2534}$^{\dagger}$$^*$ & \textbf{0.5844}$^{\dagger}$ & \textbf{0.7333}$^{\dagger}$$^*$ & \textbf{0.2681}$^{\dagger}$$^*$ & \textbf{0.2932}$^{\dagger}$$^*$ & \textbf{0.1247}$^{\dagger}$ & \textbf{0.1546}$^{\dagger}$$^*$ & \textbf{0.5064}$^{\dagger}$ & \textbf{0.6304}$^{\dagger}$\\ \bottomrule
\end{tabular}}
\label{tab:main}
\vspace{-0.3cm}
\end{table*}

\begin{table*}[t]
\caption{Retrieval performance comparison with various enhancement methods.
* indicates a significant difference (paired t-test, $p<0.05$) from the best baseline (i.e., the combination of the best existing query generation and enhancement methods). }
\renewcommand{\arraystretch}{0.9}
\resizebox{\linewidth}{!}{
\begin{tabular}{llllllllllllll}\toprule
\multicolumn{1}{c}{\multirow{2}{*}{\parbox{1.5cm}{\textbf{Query} \\ \textbf{generation}}}} & \multicolumn{1}{l}{\multirow{2}{*}{\parbox{1.5cm}{\textbf{Retrieval} \\ \textbf{enhancement}}}} & \multicolumn{6}{c}{\textbf{CSFCube}} & \multicolumn{6}{c}{\textbf{DORIS-MAE}} \\ \cmidrule(lr){3-8} \cmidrule(lr){9-14}
 &  & \textbf{N@10} & \textbf{N@20} & \textbf{M@10} & \textbf{M@20} & \textbf{R@50} & \textbf{R@100} & \textbf{N@10} & \textbf{N@20} & \textbf{M@10} & \textbf{M@20} & \textbf{R@50} & \textbf{R@100} \\ \midrule
\multirow{3}{*}{\parbox{1.5cm}{\centering Pair-wise \\generation}} & Retriever & 0.3418 & 0.3686 & 0.1522 & 0.1971 & 0.5961 & 0.7225 & 0.2541 & 0.2753 & 0.1178 & 0.1445 & 0.4809 & 0.5947 \\
 & w/ GRF & 0.3401 & 0.3713 & 0.1540 & 0.2008 & 0.5778 & 0.7151 & 0.2535 & 0.2753 & 0.1147 & 0.1416 & 0.4832 & 0.6159 \\
 & w/ ToTER & \textbf{0.3745} & \textbf{0.4072} & \textbf{0.1719} & \textbf{0.2267} & \textbf{0.6352} & \textbf{0.7606} & \textbf{0.2932} & \textbf{0.3138} & \textbf{0.1381} & \textbf{0.1680} & \textbf{0.5361} & \textbf{0.6579} \\ \midrule
\multirow{4}{*}{\parbox{1.7cm}{\centering w/ CCQGen\\(ours)}} & Retriever & 0.3670 & 0.4063 & 0.1656 & 0.2228 & 0.6362 & 0.7526 & 0.2783 & 0.2943 & 0.1308 & 0.1577 & 0.5089 & 0.6331 \\
 & w/ GRF & 0.3741 & 0.4071 & 0.1715 & 0.2272 & 0.6288 & 0.7490 & 0.2709 & 0.2925 & 0.1262 & 0.1542 & 0.5138 & 0.6384 \\
 & w/ ToTER & 0.4023 & 0.4205 & 0.1844 & 0.2403 & \textbf{0.6441} & 0.7698 & 0.2965 & 0.3159 & 0.1394 & 0.1697 & 0.5391 & 0.6635 \\
 & w/ \proposedtwo (ours) & \textbf{0.4244}$^*$ & \textbf{0.4359}$^*$ & \textbf{0.2029}$^*$ & \textbf{0.2530}$^*$ & 0.6412 & \textbf{0.7792}$^*$ & \textbf{0.3034}$^*$ & \textbf{0.3237}$^*$ & \textbf{0.1438}$^*$ & \textbf{0.1745}$^*$ & \textbf{0.5588}$^*$ & \textbf{0.6818}$^*$ \\ \bottomrule
\end{tabular}}
\label{tab:csr}
\vspace{-0.3cm}
\end{table*}

\subsection{Performance Comparison}
\label{sec:experimentresult}

\subsubsection{\textbf{Effectiveness of \proposed}}
\label{result:CCQGen}
Table \ref{tab:main} presents retrieval performance after fine-tuning with various query generation methods.
\proposed consistently outperforms all baselines, achieving significant improvements across various metrics with both backbone models.
We observe that GenQ underperforms compared to LLM-based methods, showing the advantages of leveraging the text generation capability of LLMs.
Also, existing methods often fail to improve the backbone model (i.e., no Fine-Tune), particularly \ctr.
As it is trained on labeled data from general domains, it already captures overall textual similarities well, making further improvements challenging.
The consistent improvements by \proposed support its efficacy in generating queries that effectively represent the scientific documents.
Notably, Promptgator\_diverse struggles to produce consistent improvements.
We observe that it often generates redundant queries covering similar aspects, despite increased diversity in their expressions (further analysis provided in \cref{result:query_analysis}).
This underscores the importance of proper control over generated content and supports the validity of our~approach.

\begin{figure}[t]
\centering
\includegraphics[height=3cm]{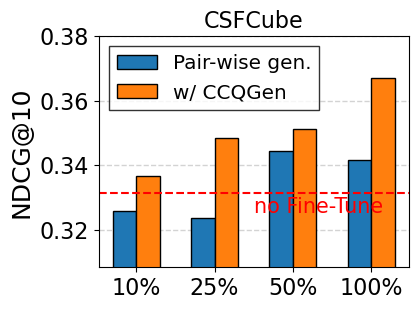}
\includegraphics[height=3cm]{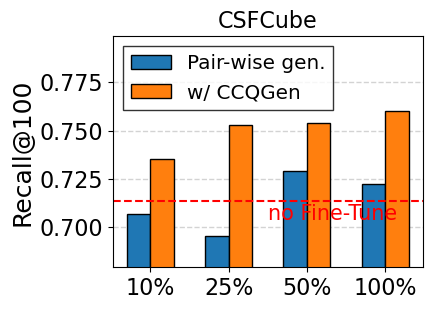}\hspace{-0.1cm}\\ \hspace{-0.3cm}
\includegraphics[height=3cm]{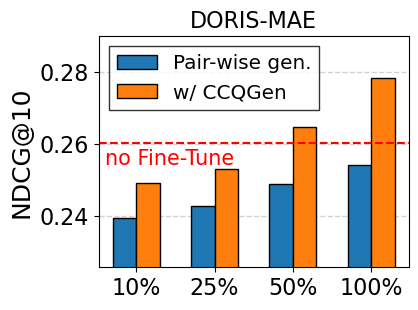}
\includegraphics[height=3cm]{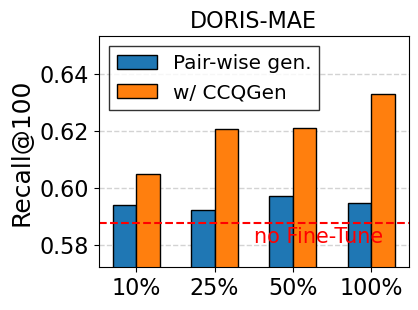}\hspace{-0.1cm}
\caption{Results with varying amounts of training data.
x\% denotes setups using a random x\% of generated queries.
}
\label{fig:amount}
\end{figure}

\smallsection{Impact of amount of training data}
In Figure \ref{fig:amount}, we further explore the retrieval performance by limiting the amount of training data, using \ctr as the backbone model.
The existing LLM-based generation method (i.e., Pair-wise gen.) shows limited performance under restricted data conditions and fails to fully benefit from an increasing volume of training data.
This supports our claim that the generated queries are often redundant and do not effectively introduce new training signals.
Conversely, \proposed consistently delivers considerable improvements, even with a limited number of queries.
\proposed guides each new query to complement the previous ones, allowing for reducing redundancy and fully leveraging the limited number of queries.

\subsubsection{\textbf{Effectiveness of \proposed with CSR}}
In \cref{subsub:method_filtering}, we introduce \proposedtwo, designed to enhance retrieval using concept information from \proposed.
This technique aligns with the ongoing research direction of enhancing retrieval by integrating additional context not directly revealed from the queries and document \cite{ToTER, mackie2023generative, BERTQE, DensePRF}.
We compare \proposedtwo with two recent methods:
(1) \textbf{GRF}~\cite{mackie2023generative} generates relevant contexts by LLMs. For a fair comparison, we generate both topics and keywords, as used in \proposed.
(2) \textbf{ToTER}~\cite{ToTER} uses the topic distributions between queries and documents, with topics provided by~the~taxonomy.
\ctr is used as~the~backbone.

Table \ref{tab:csr} presents the retrieval performance with various enhancement methods.
\proposedtwo significantly improves the retrieval performance.
Notably, the combination of proposed concept-based query generation (\proposed) and enhancement (\proposedtwo) methods achieves significant improvements over the best existing solutions (i.e., Pair-wise gen. combined with ToTER).
GRF often degrades performance because the LLM-generated contexts are not tailored to target documents;
these contexts may be related but often not covered by documents in the corpus, potentially causing discrepancies in focused aspects.
Lastly, ToTER only considers topic-level information, which may be insufficient for providing find-grained details necessary to distinguish between topically-similar documents.

\subsection{Study of \proposed}

\subsubsection{\textbf{Analysis of generated queries}}
\label{result:query_analysis}
We analyze whether \proposed indeed reduces redundancy among the queries and includes a variety of related terms.
We introduce two criteria: (1) \textbf{redundancy}, measured as the average cosine similarity of term frequency vectors of queries.\footnote{We use CountVectorizer from the SciKit-Learn library.}
A high redundancy indicates that queries tend to cover similar aspects of the document.
(2) \textbf{lexical overlap}, measured as the average BM25 score between the queries and the document.
A higher lexical overlap indicates that queries tend to reuse~terms~from~the~document.

In Table~\ref{tab:query_analysis}, the generated queries show higher lexical overlap with the document compared to the actual user queries.
This shows that the generated queries tend to use a limited range of terms already present in the document, whereas actual user queries include a broader variety of terms.
With the ‘diverse condition’ (i.e., Promptgator\_diverse), the generated queries exhibit reduced lexical overlap and redundancy. 
However, this does not consistently lead to performance improvements.
The improved term usage often appears in common expressions, not necessarily enhancing concept coverage.
Conversely, \proposed directly guides each new query to complement the previous ones.
Also, \proposed incorporate concept-related terms not explicitly mentioned in the document via enrichment step (\cref{method:enrich}).
This provides more systematic controls over the generation, leading to consistent improvements.

\subsubsection{\textbf{Effectiveness of concept coverage-based filtering}}
\begin{figure}[t]
\centering
\includegraphics[height=2.3cm]{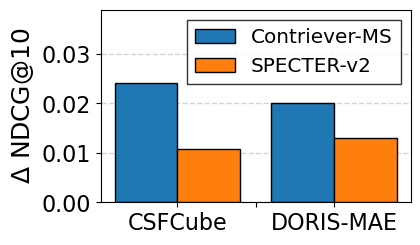}
\includegraphics[height=2.3cm]{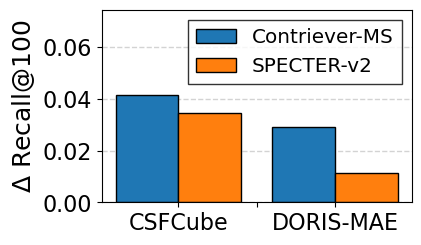}\hspace{-0.1cm}
\caption{Improvements by concept coverage-based filtering.}
\label{fig:filtering}
\end{figure}

Figure~\ref{fig:filtering} presents the improvements achieved through the filtering step, which aims to remove low-quality queries that the document does not answer (\cref{subsub:method_filtering}).
As shown in Table~\ref{tab:csr}, \proposedtwo largely enhances retrieval accuracy by incorporating concept information.
This enhanced accuracy helps to accurately measure round-trip consistency, effectively improving the effects of fine-tuning.

\begin{table}[t]
\caption{Analysis of generated queries.
(a) Statistics of queries generated by different methods.
(b) Retriever performance (\specter on NDCG@10) after fine-tuning using the queries.
The average lexical overlap of actual queries is $13.32$ for \csfcube and $20.42$ for \dorismae.}
\textbf{\csfcube}
\resizebox{1.02\linewidth}{!}{
\begin{tabular}{l l l l} \toprule
\multirow{2}{*}{\textbf{ Query generation}} & \multicolumn{2}{c}{\textbf{(a) Query statistics}} &  \multirow{2}{*}{ \parbox{1.9cm}{\centering \textbf{(b) Retriever} \\\textbf{performance}}}\\ \cmidrule(lr){2-3}
 & \textbf{redundancy} ($\downarrow$) & \textbf{lexical overlap} ($\downarrow$)& \\ \midrule
Promptgator & 0.5072 & 31.51 & 0.3766 \\
w/ diverse condition & 0.4512 (-11.0\%) & \textbf{24.05} \textbf{(-23.7\%)} & 0.3672 (-2.6\%) \\
w/ \proposed & \textbf{0.3997} \textbf{(-21.2\%)} & 24.41 (-22.5\%) & \textbf{0.4105} \textbf{(+9.0\%)} \\ \bottomrule
\end{tabular}}
\vspace{0.03cm}
\textbf{\dorismae} 
\resizebox{\linewidth}{!}{
\begin{tabular}{l l l l} \toprule
\multirow{2}{*}{\textbf{Query generation}} & \multicolumn{2}{c}{\textbf{(a) Query statistics}} &  \multirow{2}{*}{ \parbox{1.9cm}{\centering \textbf{(b) Retriever} \\\textbf{performance}}}\\ \cmidrule(lr){2-3}
 & \textbf{redundancy} ($\downarrow$) & \textbf{lexical overlap} ($\downarrow$) & \\ \midrule
Promptgator	&0.4861 & 53.58&	0.2479 \\
w/ diverse condition&	\textbf{0.3958 (-18.6\%)} & 41.56 (-22.4\%)&	0.2469 (-0.4\%) \\
w/ \proposed & 0.3993 (-17.9\%) & \textbf{40.54 (-24.3\%)}	&	\textbf{0.2634 (+6.2\%)} \\ \bottomrule
\end{tabular}}
\label{tab:query_analysis}
\vspace{-0.4cm}
\end{table}

\subsubsection{\textbf{Results with a smaller LLM}}
\label{result:Llama}
In Table~\ref{tab:Llama}, we explore the effectiveness of the proposed approach using a smaller LLM, Llama-3-8B, with \ctr as the backbone model. 
Consistent with the trends observed in Table~\ref{tab:main} and Table~\ref{tab:csr}, the proposed techniques (\proposed and \proposedtwo) consistently improve the existing method.
We expect \proposed to be effective with existing LLMs that possess a certain degree of capability. 
Since comparing different LLMs is not the focus of this work, we leave further investigation on more various LLMs and their comparison for future study.

\begin{table}[t]
\caption{Retrieval performance with Llama-3-8B. 
We report improvements over no Fine-Tune. $^*$ denotes $p < 0.05$ from paired t-test with pair-wise generation.}
\centering
\renewcommand{\arraystretch}{0.85}
\resizebox{\linewidth}{!}{
\begin{tabular}{c l lll} \toprule
\textbf{Dataset} & \textbf{Method} & \textbf{N@10} & \textbf{N@20} & \textbf{R@100} \\ \midrule
\multirow{3}{*}{CSFCube} & Pair-wise generation & +5.25\% & +0.94\% & \textcolor{red}{- 0.21\%} \\
 & w/ CCQGen & +6.55\% & +7.82\%$^*$ & +5.48\%$^*$ \\
 & w/ CCQGen + \proposedtwo & +27.92\%$^*$ & +20.09\%$^*$ & +9.01\%$^*$ \\ \midrule
\multirow{3}{*}{DORIS-MAE} & Pair-wise generation & +0.00\% & +2.92\% & +5.43\% \\
 & w/ CCQGen & +5.19\%$^*$ & +9.57\%$^*$ & +6.69\% \\
 & w/ CCQGen + \proposedtwo & +16.75\%$^*$ & +20.87\%$^*$ & +14.65\%$^*$\\ \bottomrule
\end{tabular}}
\label{tab:Llama}
\end{table}

\section{Related Work}
\label{sec:relatedwork}
We provide a detailed survey of LLM-based query generation~in~\cref{prelim:qgen}.

\smallsection{PLM-based retrieval models}
The advancement of PLMs has led to significant progress in retrieval.
Recent studies have introduced retrieval-targeted pre-training \cite{CTR, condenser}, distillation from cross-encoders \cite{AR2}, and advanced negative mining methods \cite{zhan2021optimizing, rocketqa_v1}
There is also an increasing emphasis on pre-training methods specifically designed for the scientific domain. 
In addition to pre-training on academic corpora \cite{SCIBERT}, researchers have exploited metadata associated with scientific papers. 
\cite{razdaibiedina2023miread} uses journal class, \cite{SPECTER, SCINCL} use citations, \cite{ASPIRE} uses co-citation contexts, and \cite{OAGBERT} utilizes venues, affiliations, and authors.
\cite{SPECTER2, zhang2023pre} devise multi-task learning of related tasks such as citation prediction~and~paper~classification.
Very recently, \cite{ToTER, taxoindex} leverage corpus-structured knowledge (e.g., core topics and phrases) for academic concept matching.


\smallsection{Synthetic query generation}
Earlier studies \cite{nogueira2019document, nogueira2019doc2query, liang2020embedding, ma2021zero, wang2022gpl} have employed dedicated query generation models, trained using massive document-query pairs from general domains.
Recently, there has been a shift towards replacing these generation models with LLMs \cite{dai2022promptagator, inpars, inpars2, pairwise_qgen, saad2023udapdr, sachan2022improving}, as discussed in \cref{prelim:qgen}.
On the other hand, many recent studies have focused on developing query generation tailored to specific retrieval domains.
\cite{synthetic_apple_VA} focuses on entity search for virtual assistants,
\cite{shen2022diversified} improves the diversity of queries for news article searches guided by a knowledge graph,
\cite{controllable_QGen} focuses on enhancing the retrievability of entities on online content (e.g., Podcast) platforms.
However, a dedicated method for scientific document retrieval has not been studied well in the literature.

\vspace{-0.3cm}


\section{Conclusion}
\label{sec:conclusion}
We propose \proposed framework to generate a set of queries that comprehensively cover the document concepts.
\proposed identifies concepts not sufficiently covered by previous queries, and leverages them as conditions for subsequent query generation.
This approach guides each new query to complement the previous ones, aiding in a comprehensive understanding of the document.
Extensive experiments show that \proposed significantly improves both query quality and retrieval performance, even with limited training data. 
Future work may explore its applicability across various domains. 
In particular, e-commerce \cite{SSCDR, DERRD, MvFS} presents a promising opportunity, as users often express multi-faceted needs involving desired attributes, characteristics, or specific use cases. 
We expect \proposed to better simulate user queries in such scenarios and leave further investigation for future work.

\noindent
\textbf{Acknowledgements.}
This work was supported IITP grant funded by MSIT (No.2018-0-00584, No.2019-0-01906), NRF grant funded by the MSIT (No.RS-2023-00217286, No.2020R1A2B5B03097210).
It was also in part by US DARPA INCAS Program No. HR0011-21-C0165 and BRIES Program No. HR0011-24-3-0325, National Science Foundation IIS-19-56151, the Molecule Maker Lab Institute: An AI Research Institutes program supported by NSF under Award No. 2019897, and the Institute for Geospatial Understanding through an Integrative Discovery Environment (I-GUIDE) by NSF under Award No. 2118329.

\pagebreak
\newpage
\clearpage

\bibliographystyle{ACM-Reference-Format}         
\bibliography{main}

\end{document}